\documentclass[a4paper,fleqn]{cas-sc}

\usepackage[numbers]{natbib}

\def\tsc#1{\csdef{#1}{\textsc{\lowercase{#1}}\xspace}}
\tsc{WGM}
\tsc{QE}
\begin{document}
\let\WriteBookmarks\relax
\def\floatpagepagefraction{1}
\def\textpagefraction{.001}

% Short title
\shorttitle{Self-organized step-terrace structure in SrRuO$_3$ thin films}  

% Short author
\shortauthors{R. Arakawa et al.}

% Main title of the paper
\title[mode = title]{Self-organized formation of step-terrace structure 
in \texorpdfstring{SrRuO$_3$}{SrRuO3} thin films grown on 
mixed-terminated \texorpdfstring{SrTiO$_3$}{SrTiO3} (100) substrates} 

%==1st author===============
\author[1]{Ryotaro Arakawa}
\cormark[1]
\credit{Conceptualization, Methodology, Writing -- original
draft, Writing -- review \& editing, Data curation, Formal analysis, Visualization, Validation, Software, Investigation}
\affiliation[1]{organization={Department of Physics, Nagoya University},
            city={Nagoya},
            postcode={464-8602}, 
            state={Aichi},
            country={Japan}}
%\ead{arakawa.ryotaro.u3@s.mail.nagoya-u.ac.jp}
%==2nd author===============
\author[2]{Sachio Komori}
\cormark[0]
\credit{Supervision, Writing -- review \& editing}
\affiliation[2]{organization={Department of Physics and Electronics, Osaka Metropolitan University},
            city={Sakai},
            postcode={599-8531}, 
            state={Osaka},
            country={Japan}}
%==3rd author===============
\author[1]{Kotaro Tomita}
\cormark[0]
\credit{Investigation, Methodology}
%==4th author===============
\author[1]{Shunsei Komori}
\cormark[0]
\credit{Investigation, Validation}
%==5th author===============
\author[3]{Masaaki A. Tanaka}
\cormark[0]
\credit{Methodology}
\affiliation[3]{organization={Department of Physical Science and Engineering, Nagoya Institute of Technology},
                city={Nagoya},
                postcode={466-8555},
                state={Aichi},
                country={Japan}}
%==6th author===============
\author[1]{Tomoyasu Taniyama}[orcid=0000-0003-3683-5416]
\cormark[2]
\credit{Supervision, Resources, Funding acquisition, Project administration, Writing -- review \& editing}
%\ead{taniyama.tomo@nagoya-u.jp}
%===========================
\cortext[1]{Corresponding author}

\begin{abstract}
Surface morphology of the substrate and bottom layers plays a critical role in the epitaxial growth of oxide thin films. Here, we report on the self-organized formation of a step-terrace structure in SrRuO$_3$ (SRO) thin films grown using pulsed laser deposition on mixed-terminated SrTiO$_3$ (100) substrates without any prior surface treatment. Atomic force microscopy observations reveal that SRO films initially grow in a three-dimensional island mode and subsequently undergo a transition to a step-flow growth mode through island coalescence as the film thickness increases, resulting in a well-defined step-terrace morphology with a step height consistent with the SRO unit-cell parameter. The average terrace width of the self-organized structure can be systematically tuned by varying the substrate temperature and the target-substrate distance, which we attribute to changes in the critical island radius that governs the nucleation behavior. To demonstrate the utility of this self-organized morphology, we show that BiFeO$_3$ thin films grown on SRO films with such a step-terrace structure exhibit improved surface flatness and crystalline quality compared to those grown directly on bare SrTiO$_3$ substrates. These findings provide a clear understanding of the mechanism of thickness-driven growth-mode transitions in perovskite oxide thin films under various growth conditions.
\end{abstract}
%=Graphical abstract=========================================================

%============================================================================


\maketitle

%=Introduction==========================================
\section{Introduction}\label{sec:introduction}
SrRuO$_3$ (SRO) thin films are a widely studied perovskite material, well known for their $4d$ itinerant metallic conductivity and are commonly used as bottom electrode materials in oxide electronics \cite{chang2009fundamental,cuoco2022materials}. It is generally crucial to control the surface morphology of the bottom electrode layer to achieve high-quality epitaxial growth of subsequent layers, and SRO films are no exception \cite{wang2024surface}. In this regard, substrates with a well-defined step-terrace structure are widely employed for the growth of high-quality epitaxial films.

The surface morphology of SRO thin films, however, is known to exhibit complex growth behaviors depending on the substrate and growth conditions. For instance, SRO thin films grown on vicinal SrTiO$_3$ (STO) (100) substrates which possess a well-defined step-terrace structure exhibit a finger-like surface morphology in which a complete step-terrace structure fails to form at low film thicknesses \cite{sanchez2004critical}, while in other cases the surface roughness decreases with increasing film thickness, suggesting a self-healing behavior \cite{rubi2009growth}. More remarkably, SRO thin films grown on exactly-oriented STO (100) substrates entirely lacking an initial step-terrace structure have been reported to spontaneously develop a step-terrace morphology \cite{zhu2006epitaxial,jiang2017epitaxial}---a phenomenon that has not typically been observed in conventional thin-film systems, where step-terrace structures form exclusively through the diffusion of adatoms along pre-existing step edges on single-terminated substrates \cite{hong2005persistent,lee2004thermal,schraknepper2015complex}. Despite these observations, the mechanism underlying this spontaneous step-terrace formation remains poorly understood.

Given that SRO is one of the most widely used bottom electrode materials in perovskite oxide heterostructures, understanding its growth mechanism and achieving flat, well-defined surface morphologies is of both fundamental and practical importance. In this study, we systematically investigate the surface morphology of SRO thin films grown on mixed-terminated STO (100) substrates without any initial step-terrace structure, with the aim of elucidating the growth mechanism responsible for the spontaneous step-terrace formation. We find that the SRO films initially grow in a three-dimensional (3D) island growth mode and subsequently transition to a step-flow growth mode through the coalescence of 3D islands as the film thickness increases, resulting in a self-organized step-terrace structure with relatively well-defined step edges. Although a thickness-driven transition from 3D island growth to step-flow growth has been reported in semiconductor heteroepitaxial systems, such a systematic thickness-dependent morphological evolution has not been reported in perovskite oxide thin films. 

We further show that the terrace width of the self-organized step-terrace structure is strongly influenced by the critical island radius, which can be systematically controlled by the target-substrate distance and the substrate temperature. To demonstrate the potential utility of this morphology, we show that BiFeO$_3$ (BFO) thin films grown on SRO films with a self-organized step-terrace structure exhibit improved surface flatness and crystalline quality compared to those grown directly on bare STO (100) substrates. These findings provide new insights into the growth mechanism of self-organized step-terrace structures in perovskite oxide thin films.

%=Methods===============================================
\section{Experimental Methods}\label{sec:methods}
Epitaxial SRO thin films were grown on as-received STO (100) substrates with a miscut angle of less than $0.2^\circ$ by pulsed laser deposition (PLD) using a Nd:YAG laser with a wavelength of 266~nm. The STO substrates were used without any prior surface treatment and are therefore mixed-terminated with no step-terrace structure. It is noted that droplets are commonly formed when a Nd:YAG laser is used, owing to spatially inhomogeneous ablation \cite{chaluvadi2021pulsed}. In this study, a 3D-shadow mask was introduced between the target and the substrate to prevent the deposition of droplets \cite{tachiki2000improved}, successfully yielding atomically flat, highly crystalline SRO films free of droplets.

The growth conditions for SRO thin films are as follows: substrate temperature $T = 700$ and $725^\circ$C, oxygen pressure $P_{\text{O}_2} = 80$--$120$~mTorr, laser fluence $F = 6$~J/cm$^2$, repetition rate $f = 10$~Hz, target-substrate distance $d = 4$--$4.75$~cm, and deposition time $t = 30$--$120$~min, yielding a film thickness of approximately 34~nm at $t = 120$~min. To investigate the effect of the self-organized step-terrace structure on the growth of subsequent layers, BFO thin films were also grown on the SRO films under the following conditions: $T = 650^\circ$C, $P_{\text{O}_2} = 100$~mTorr, $F = 1.2$~J/cm$^2$, $f = 10$~Hz, $d = 4.5$~cm, and $t = 180$~min, yielding a film thickness of approximately 33~nm. All samples were cooled to room temperature over approximately 2~hours under $P_{\text{O}_2} = 600$~Torr immediately after growth.

The surface morphology of the films was observed by atomic force microscopy (AFM) in non-contact mode using a Park Systems NX7. The crystal structure of the films was characterized by reciprocal space mapping (RSM) using a Rigaku SmartLab X-ray diffractometer (XRD).

%=Results and Discussion================================
\section{Results and Discussion}\label{sec:results_and_discussion}
Figure~\ref{fig:AFM_STO} shows the surface morphology of the typical STO (100) substrates used in this study. The surface of the as-received STO substrate in Fig.~\ref{fig:AFM_STO}(a) is flat with no step-terrace structure. We also annealed an STO substrate at $700^\circ$C for 120~min under an oxygen pressure of 100~mTorr—the same conditions as those used for SRO film growth—to examine whether a step-terrace structure can be induced by annealing at the growth temperature. The annealed STO substrate does not exhibit a clear step-terrace structure (Fig.~\ref{fig:AFM_STO}(b)), which is consistent with previous reports showing that in-situ annealing of STO substrates produces large terraces with straight steps only at temperatures above $900$--$1100^\circ$C \cite{jager2018independence}. These results indicate that the self-organized formation of the step-terrace structure in SRO thin films cannot be attributed to the surface morphology of the STO substrate.

SRO films deposited on as-received STO substrates without a step-terrace structure form a well-defined step-terrace structure under growth conditions of $T = 725^\circ$C, $P_{\text{O}_2} = 100$~mTorr, and $t = 120$~min, as shown in Fig.~\ref{fig:AFM_SRO_Lineprofile}. The line profile taken along the white line in Fig.~\ref{fig:AFM_SRO_Lineprofile}(a) reveals a step-terrace structure with an average terrace width of approximately 220~nm and a step height of approximately 0.4~nm, comparable to the pseudocubic unit-cell parameter of SRO (${\sim}0.393$~nm) \cite{koster2012structure}.
This step-terrace morphology also forms when SRO is deposited without the 3D-shadow mask under different growth conditions (see Fig.~S1); however, it is sensitive to the oxygen pressure during deposition, and a slight change in $P_{\text{O}_2}$ leads to the considerable degradation of the step-terrace structure and the occurrence of multiple surface defects (see Fig.~S2).

To investigate the growth mechanism of the self-organized step-terrace structure, we deposited SRO films for varying growth times (thickness) and observed the evolution of the surface morphology. As shown in Fig.~\ref{fig:AFM_SRO_time_dependence}, SRO films grown for 30~min exhibit a 3D island growth mode in which the surface is dominated by a high density of islands. As the growth time increases, the islands grow into elongated oval shapes roughly aligned along the [110] direction and coalesce with each other (60~min), leading to the formation of a step-terrace structure with some large residual pits (90~min) at the junctions where adjacent islands merge. These pits are eventually filled by subsequent growth, and the step-terrace structure becomes more well defined (120~min), at which the film thickness reaches approximately 34~nm.
These results demonstrate a growth-mode transition in SRO thin films on STO (100) substrates from a 3D island mode to a step-flow mode, which has been reported in semiconductor heteroepitaxial 
systems such as GaN on sapphire~\cite{takemura2024metal} and ZnSe on 
GaAs~\cite{zhang2001thickness} but not in perovskite oxide thin films.
This self-organized formation of a step-terrace structure is fundamentally different from the finger-like morphology previously reported for SRO films on as-received STO (100) substrates \cite{sanchez2006growth} in which SRO does not form complete step-terrace structures. Step-terrace structures are typically formed either by the coalescence of two-dimensional (2D) islands on a terrace referred to as layer-by-layer growth \cite{chang2016direct} or by the diffusion of adatoms to step edges, referred to as step-flow growth \cite{HandbookCrystalGrowthIII_A}. In either case, a single-terminated substrate with a well-defined step-terrace structure is required for the overlying film to develop a step-terrace morphology.

It should also be noted that the average terrace width $L_\text{ave}$ of the self-organized step-terrace structure is larger for SRO films grown at $725^\circ$C (Fig.~\ref{fig:AFM_SRO_Lineprofile}(a)) than for those grown at $700^\circ$C (Fig.~\ref{fig:AFM_SRO_time_dependence}(d)), approximately 220~nm and 120~nm, respectively. This difference can be attributed to the critical island radius $R_\text{c}$, which defines the minimum island size that can remain stable against the re-evaporation of adatoms \cite{amar1995critical} and is governed by thermal conditions such as the substrate temperature. When $R_\text{c}$ is smaller than the island separation, new islands tend to nucleate on top of existing 2D islands, giving rise to 3D island structures \cite{tersoff1994critical}. For SRO films grown at $725^\circ$C (Fig.~\ref{fig:Schematics_Island_growth}(a)), the elevated substrate temperature suppresses the nucleation of small, unstable islands, resulting in a larger $R_\text{c}$, larger islands, and consequently a larger $L_\text{ave}$ after coalescence. In contrast, a lower substrate temperature of $700^\circ$C (Fig.~\ref{fig:Schematics_Island_growth}(b)) favors a smaller $R_\text{c}$, leading to finer islands and a smaller $L_\text{ave}$.

It has been theoretically shown for multilayer growth~\cite{Markov2016}, in which 3D islands composed of bottom-2D island with a radius $\rho_1$ and top-2D island with a radius $\rho_2$ grow simultaneously ($\rho_1>\rho_2$), that $\rho_1^2 - \rho_2^2$ is proportional to $1/N_\text{s}$, where $N_\text{s}$ denotes the saturation island density, which is inversely related to $R_\text{c}$. In this configuration, the quantity $\rho_1 - \rho_2$ represents the lateral distance between the edges of the upper and lower islands, corresponding to the step-terrace width that self-organizes in this study. A higher substrate temperature increases $R_\text{c}$, which in turn decreases $N_\text{s}$ and leads to an increase in $\rho_1 - \rho_2$, consistent with the larger $L_\text{ave}$ observed at $725^\circ$C.

This growth picture is further supported by the dependence of $L_\text{ave}$ on the target-substrate distance $d$, as shown in Fig.~\ref{fig:AFM_SRO_dTS_dependence}. SRO films grown with $d = 4.0$~cm exhibit a very small $L_\text{ave}$ of approximately 73~nm (Fig.~\ref{fig:AFM_SRO_dTS_dependence}(a)), accompanied by some large square-shaped pits, whereas those grown with $d = 4.75$~cm exhibit flat terraces with a larger $L_\text{ave}$ of approximately 110~nm (Fig.~\ref{fig:AFM_SRO_dTS_dependence}(d)). Notably, $L_\text{ave}$ increases linearly with increasing $d$, as shown in Fig.~\ref{fig:AFM_SRO_dTS_dependence}(e). This difference in $L_\text{ave}$ can be attributed to the kinetic energy of the incident atoms, which is governed by $d$. At smaller $d$ (Fig.~\ref{fig:Schematics_Island_growth_dTS}(a)), the incident atoms possess higher kinetic energy, leading to local surface damage such as pit formation through energetic collisions with the substrate. These pits serve as preferential nucleation sites, resulting in a higher nucleation density, smaller islands, and a reduced $R_\text{c}$, ultimately yielding a smaller $L_\text{ave}$. At larger $d$ (Fig.~\ref{fig:Schematics_Island_growth_dTS}(b)), the lower kinetic energy suppresses pit formation, allowing islands to grow more stably and increasing both $R_\text{c}$ and $L_\text{ave}$. This interpretation agrees with kinetic Monte Carlo simulations of PLD film growth, which show that energetic collisions can split unstable small islands into single adatoms, thereby increasing nucleation density and decreasing the average island size~\cite{zhang2006simulation}.

We also investigated the effect of the self-organized step-terrace structure in SRO thin films on the growth of subsequent layers by depositing BFO thin films with a thickness of approximately 33~nm. BFO thin films grown directly on as-received STO substrates without a step-terrace structure undergo Volmer--Weber growth, yielding a rough surface dominated by 3D islands with a root-mean-square roughness of approximately 5.2~nm (Fig.~\ref{fig:AFM_RSM_BFO}(a)). Since the surface morphology of BFO films is highly sensitive to the surface structure of the underlying substrate \cite{jang2009domain,zhang2024manipulation}, BFO films grown on mixed-terminated STO (100) substrates typically exhibit an inhomogeneous surface morphology and are prone to the formation of impurity phases such as Bi$_2$O$_3$ or Fe$_2$O$_3$ \cite{toupet2009growth}.

In contrast, BFO thin films grown on SRO films with a self-organized step-terrace structure—specifically those grown under the conditions of Fig.~\ref{fig:AFM_SRO_dTS_dependence}(b)—exhibit a significantly smaller root-mean-square roughness of approximately 0.56~nm (Fig.~\ref{fig:AFM_RSM_BFO}(b)). Although the growth mode is not a perfect step-flow but rather step-bunching, in which multiple steps coalesce, further optimization of the growth conditions could enable true step-flow growth of BFO films. The crystal structure of these BFO films was also evaluated by RSM around the (103) STO Bragg peaks (Figs.~\ref{fig:AFM_RSM_BFO}(c) and (d)), with all reciprocal-space coordinates normalized to those of STO ($= 1/0.3905$~nm$^{-1}$)~\cite{chu2007domain}. In both samples, the BFO films exhibit a single RSM peak, confirming $c$-axis-aligned growth without any impurity phases. The BFO peak is sharper in the BFO/SRO/STO sample than in the BFO/STO sample, indicating improved crystalline quality owing to the self-organized step-terrace structure of the SRO underlayer. These results demonstrate that the self-organized step-terrace morphology effectively reduces surface roughness and enhances the crystalline quality of subsequently deposited films. The ferroelectric response of BFO/SRO/STO was further evaluated by piezoelectric force microscopy, confirming a clear ferroelectric domain structure (see Fig.~S3).

%=Conclusion============================================
\section{Conclusion}\label{sec:conclusion}
In conclusion, we have elucidated the growth mechanism responsible for the self-organized formation of step-terrace structures in SrRuO$_3$ films grown on mixed-terminated STO (100) substrates without any initial step-terrace structure. Systematic thickness-dependent observations reveal a clear growth-mode transition from 3D island growth to step-flow growth as the film thickness increases, driven by the coalescence of 3D islands. The terrace width of the resulting self-organized step-terrace structure can be tuned by adjusting the substrate temperature and the target-substrate distance, which we attribute to the corresponding changes in the critical island radius governing the nucleation behavior. These findings provide a clear understanding of the mechanism of thickness-driven growth-mode transitions in perovskite oxide thin films under various growth conditions. As a further demonstration, ferroelectric BiFeO$_3$ films grown on SRO films with this self-organized step-terrace structure exhibit improved surface morphology and crystallinity compared to those grown directly on bare STO substrates, suggesting that this morphology may also serve as a useful template for subsequent epitaxial growth.

%=Appendix============================================
\appendix
\section*{Appendix A: Supplementary data}
The data that support the findings of this study are available from the corresponding author upon reasonable request.

%=======================================================
\printcredits
\section*{Declaration of competing interest}
The authors declare no competing financial interests.

\section*{Declaration of generative AI use in scientific writing}
The authors used generative AI tools (Microsoft Copilot and Claude) only for language refinement. All scientific content, data interpretation, and conclusions were generated by the authors.

\section*{Acknowledgments}
This work was supported in part by JSPS KAKENHI Grant Nos. JP23KK0086, JP24H00380, JP24K21732, JSPS International Joint Research Program (JRP-LEAD with UKRI) No. JPJSJRP20241705, JST FOREST Grant No. JPMJFR212V.

\section*{Data availability}
The data that support the findings of this study are available from the corresponding author upon reasonable request.

\bibliographystyle{model1-num-names}
\bibliography{SRO_arxiv}

@string {prl    = "Phys. Rev. Lett."}

@string {jmmm   = "J. Magn. Magn. Mater."}

@string {apl    = "Appl. Phys. Lett."}

@string {prm    = "Phys. Rev. Mater."}

@string {aplm   = "APL Mater."}

@string {tsf    = "Thin Solid Films"}

@string {as     = "Appl. Sci."}

@string {jpm    = "J. Phys.: Mater."}

@string {rmp    = "Rev. Mod. Phys."}

@string {pccp   = "Phys. Chem. Chem. Phys."}

@string {acsn   = "ACS Nano"}

@string {ass    = "Appl. Surf. Sci."}

@string {mc     = "Mater. Charact."}

@string {pssc   = "Prog. Solid State Chem."}

@string {am     = "Adv. Mater."}

@string {eej    = "Electr. Eng. Jpn."}

@string {jcg    = "J. Cryst. Growth"}

@string{pssb = "Phys. Status Solidi B"}

@article{schraknepper2015complex,
  title={Complex behaviour of vacancy point-defects in {SrRuO$_3$} thin films},
  author={Schraknepper, Henning and B{\"a}umer, Christoph and Dittmann, Regina and De Souza, Roger A},
  journal=pccp,
  volume={17},
  number={2},
  pages={1060--1069},
  year={2015},
  publisher={Royal Society of Chemistry}
}

@article{chang2016direct,
  title={Direct nanoscale analysis of temperature-resolved growth behaviors of ultrathin perovskites on {SrTiO$_3$}},
  author={Chang, Young Jun and Phark, Soo-hyon},
  journal=acsn,
  volume={10},
  number={5},
  pages={5383--5390},
  year={2016},
  publisher={ACS Publications}
}

@article{sanchez2004critical,
  title={Critical effects of substrate terraces and steps morphology on the growth mode of epitaxial {SrRuO$_3$} films},
  author={S{\'a}nchez, F and Herranz, G and Infante, IC and Fontcuberta, J and Garc{\'\i}a-Cuenca, MV and Ferrater, C and Varela, M},
  journal=apl,
  volume={85},
  number={11},
  pages={1981--1983},
  year={2004},
  publisher={AIP Publishing}
}

@article{lee2004thermal,
  title={Thermal stability of epitaxial {SrRuO$_3$} films as a function of oxygen pressure},
  author={Lee, Ho Nyung and Christen, Hans M and Chisholm, Matthew F and Rouleau, Christopher M and Lowndes, Douglas H},
  journal=apl,
  volume={84},
  number={20},
  pages={4107--4109},
  year={2004},
  publisher={American Institute of Physics}
}

@article{zhang2006simulation,
  title={Simulation of island aggregation influenced by substrate temperature, incidence kinetic energy and intensity in pulsed laser deposition},
  author={Zhang, Duanming and Guan, Li and Li, Zhihua and Pan, Guijun and Tan, Xinyu and Li, Li},
  journal=ass,
  volume={253},
  number={2},
  pages={874--880},
  year={2006},
  publisher={Elsevier}
}

@article{chu2007domain,
  title={Domain control in multiferroic {BiFeO$_3$} through substrate vicinality},
  author={Chu, Ying-Hao and Cruz, Maria P and Yang, Chan-Ho and Martin, Lane W and Yang, Pei-Ling and Zhang, Jing-Xian and Lee, Kilho and Yu, Pu and Chen, Long Qing and Ramesh, Ramamoorthy},
  journal=am,
  volume={19},
  number={18},
  pages={2662},
  year={2007},
  publisher={Wiley-VCH}
}

@article{tersoff1994critical,
  title={Critical island size for layer-by-layer growth},
  author={Tersoff, J and Van Der Gon, AW Denier and Tromp, RM},
  journal=prl,
  volume={72},
  number={2},
  pages={266},
  year={1994},
  publisher={APS}
}

@article{koster2012structure,
  title={Structure, physical properties, and applications of {SrRuO$_3$} thin films},
  author={Koster, Gertjan and Klein, Lior and Siemons, Wolter and Rijnders, Guus and Dodge, J Steven and Eom, Chang-Beom and Blank, Dave HA and Beasley, Malcolm R},
  journal=rmp,
  volume={84},
  number={1},
  pages={253--298},
  year={2012},
  publisher={APS}
}

@article{jager2018independence,
  title={Independence of surface morphology and reconstruction during the thermal preparation of perovskite oxide surfaces},
  author={J{\"a}ger, Maren and Teker, Ali and Mannhart, Jochen and Braun, Wolfgang},
  journal=apl,
  volume={112},
  number={11},
  pages={111601},
  year={2018},
  publisher={AIP Publishing}
}

@article{tachiki2000improved,
  title={An improved laser ablation method using a shadow mask (eclipse method)},
  author={Tachiki, Minoru and Kobayashi, Takeshi},
  journal=eej,
  volume={130},
  number={1},
  pages={88--94},
  year={2000},
  publisher={Wiley Online Library}
}

@article{amar1995critical,
  title={Critical cluster size: Island morphology and size distribution in submonolayer epitaxial growth},
  author={Amar, Jacques G and Family, Fereydoon},
  journal=prl,
  volume={74},
  number={11},
  pages={2066},
  year={1995},
  publisher={APS}
}

@book{Markov2016,
  author    = {Ivan V. Markov},
  title     = {Crystal Growth for Beginners: Fundamentals of Nucleation, Crystal Growth and Epitaxy},
  edition   = {3rd},
  year      = {2016},
  publisher = {World Scientific Publishing Co. Pte. Ltd.},
  address   = {Hackensack, New Jersey and Singapore},
  isbn      = {9789813143425}
}

@article{jiang2017epitaxial,
  title={Epitaxial growth of {BiFeO$_3$} films on {SrRuO$_3$}/{SrTiO$_3$}},
  author={Jiang, Zhen-Zheng and Guan, Zhao and Yang, Nan and Xiang, Ping-Hua and Qi, Rui-Juan and Huang, Rong and Yang, Ping-Xiong and Zhong, Ni and Duan, Chun-Gang},
  journal=mc,
  volume={131},
  pages={217--223},
  year={2017},
  publisher={Elsevier}
}

@article{sanchez2006growth,
  title={Growth modes and self-organization in the epitaxy of ferromagnetic {SrRuO$_3$} on {SrTiO$_3$} (001)},
  author={S{\'a}nchez, F and Herranz, G and Infante, IC and Ferrater, C and Garcia-Cuenca, MV and Varela, M and Fontcuberta, J},
  journal=pssc,
  volume={34},
  number={2-4},
  pages={213--221},
  year={2006},
  publisher={Elsevier}
}

@article{toupet2009growth,
  title={Growth and thermal stability of epitaxial {BiFeO$_3$} thin films},
  author={Toupet, H and Le Marrec, F and Holc, J and Kosec, M and Vilarhino, P and Karkut, MG},
  journal=jmmm,
  volume={321},
  number={11},
  pages={1702--1705},
  year={2009},
  publisher={Elsevier}
}

@article{rubi2009growth,
  title={Growth of flat {SrRuO$_3$} (111) thin films suitable as bottom electrodes in heterostructures},
  author={Rubi, D and Vlooswijk, AHG and Noheda, Beatriz},
  journal=tsf,
  volume={517},
  number={6},
  pages={1904--1907},
  year={2009},
  publisher={Elsevier}
}

@article{zhang2024manipulation,
  title={Manipulation of {BiFeO$_3$} nanostructure by substrate terrace morphology},
  author={Zhang, Tong and Li, Junhong and Zhao, Mi and Wu, Liang and Chen, Qingming and Ma, Ji and Yi, Jianhong},
  journal=ass,
  volume={648},
  pages={159088},
  year={2024},
  publisher={Elsevier}
}

@article{jang2009domain,
  title={Domain engineering for enhanced ferroelectric properties of epitaxial (001) {BiFeO$_3$} thin films},
  author={Jang, Ho Won and Ortiz, Daniel and Baek, Seung-Hyub and Folkman, Chad M and Das, Rasmi R and Shafer, Padraic and Chen, Yanbin and Nelson, Christofer T and Pan, Xiaoqing and Ramesh, Ramamoorthy and others},
  journal=am,
  volume={21},
  number={7},
  pages={817--823},
  year={2009},
  publisher={WILEY-VCH Verlag Weinheim}
}

@article{cuoco2022materials,
  title={Materials challenges for {SrRuO$_3$}: From conventional to quantum electronics},
  author={Cuoco, Mario and Di Bernardo, Angelo},
  journal=aplm,
  volume={10},
  number={9},
  pages={090902},
  year={2022},
  publisher={AIP Publishing}
}

@article{wang2024surface,
  title={Surface termination effect of {SrTiO$_3$} substrate on ultrathin {SrRuO$_3$}},
  author={Wang, Huiyu and Wang, Zhen and Ali, Zeeshan and Wang, Enling and Saghayezhian, Mohammad and Guo, Jiandong and Zhu, Yimei and Tao, Jing and Zhang, Jiandi},
  journal=prm,
  volume={8},
  number={1},
  pages={013605},
  year={2024},
  publisher={APS}
}

@article{hong2005persistent,
  title={Persistent step-flow growth of strained films on vicinal substrates},
  author={Hong, Wei and Lee, Ho Nyung and Yoon, Mina and Christen, Hans M and Lowndes, Douglas H and Suo, Zhigang and Zhang, Zhenyu},
  journal=prl,
  volume={95},
  number={9},
  pages={095501},
  year={2005},
  publisher={APS}
}

@article{chang2009fundamental,
  title={Fundamental thickness limit of itinerant ferromagnetic {SrRuO$_3$} thin films},
  author={Chang, Young Jun and Kim, Choong H and Phark, S-H and Kim, YS and Yu, J and Noh, TW},
  journal=prl,
  volume={103},
  number={5},
  pages={057201},
  year={2009},
  publisher={APS}
}

@article{chaluvadi2021pulsed,
  title={Pulsed laser deposition of oxide and metallic thin films by means of {Nd:YAG} laser source operating at its 1st harmonics: recent approaches and advances},
  author={Chaluvadi, SK and Mondal, D and Bigi, C and Knez, D and Rajak, P and Ciancio, R and Fujii, J and Panaccione, G and Vobornik, I and Rossi, G and others},
  journal=jpm,
  volume={4},
  number={3},
  pages={032001},
  year={2021},
  publisher={IOP Publishing}
}

@article{zhu2006epitaxial,
  title={Epitaxial growth and ferroelectric properties of {Pb(Zr$_{0.52}$Ti$_{0.48}$)O$_3$}/{SrRuO$_3$} heterostructures on exact {SrTiO$_3$} (001) substrates},
  author={Zhu, TJ and Lu, L and Zhao, XB and Ji, ZG and Ma, J},
  journal=jcg,
  volume={291},
  number={2},
  pages={385--389},
  year={2006},
  publisher={Elsevier}
}

@book{HandbookCrystalGrowthIII_A,
  title     = {Handbook of Crystal Growth: Thin Films and Epitaxy: Basic Techniques},
  volume    = {III, Part A},
  edition   = {Second},
  editor    = {Thomas F. Kuech},
  editor-in-chief = {Tatau Nishinga},
  publisher = {Elsevier},
  year      = {2015},
  isbn      = {9780444633062},
}

@article{takemura2024metal,
  title={Metal--Organic Vapor Phase Epitaxy of High-Quality {GaN} on {Al}-Pretreated Sapphire Substrates Without Using Low-Temperature Buffer Layers},
  author={Takemura, Kodai and Fukui, Takato and Matsuda, Yoshinobu and Funato, Mitsuru and Kawakami, Yoichi},
  journal=pssb,
  volume={261},
  number={11},
  pages={2400043},
  year={2024},
  publisher={Wiley Online Library}
}

@article{zhang2001thickness,
  title={Thickness dependent surface morphologies and luminescent properties of {ZnSe} epilayers grown on (001) {GaAs} by metalorganic chemical vapor phase deposition},
  author={Zhang, X. B. and Ha, K. L. and Hark, S. K.},
  journal=jcg,
  volume={223},
  number={4},
  pages={528--534},
  year={2001},
  publisher={Elsevier}
}

%=Figure=============================
\begin{figure}[h]
  \centering
  \includegraphics[width=5.5cm]{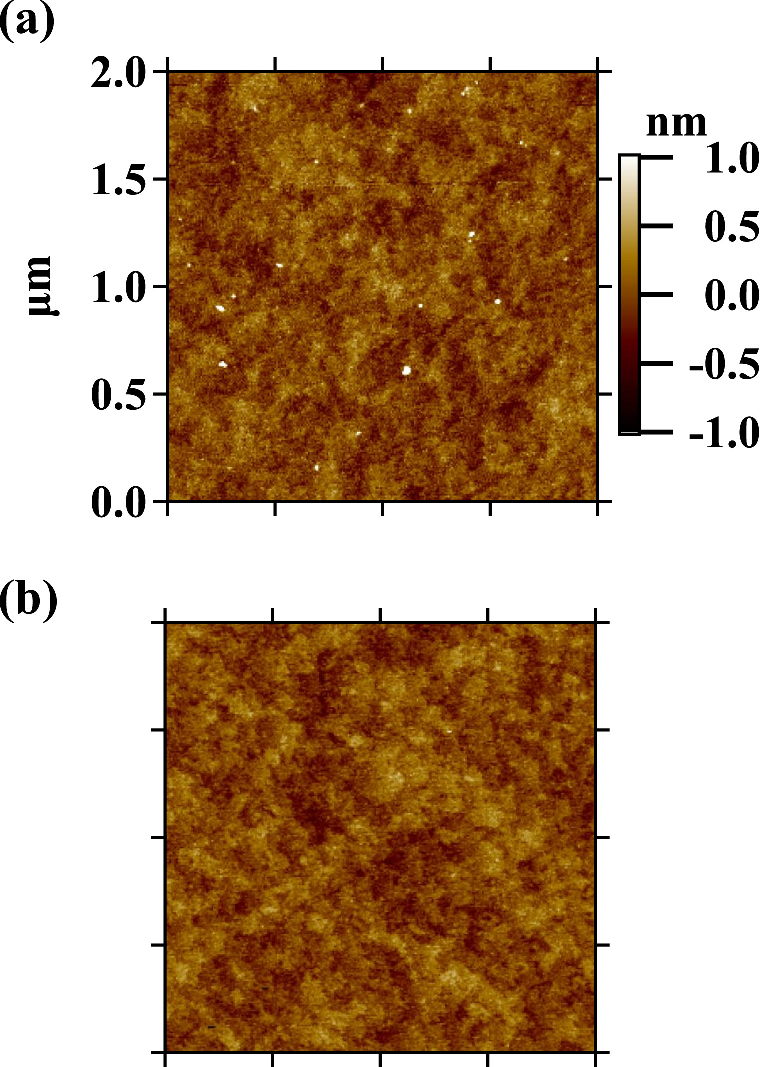}
  \caption{Surface morphology of (a)~as-received and (b)~PLD-annealed ($T = 700^\circ$C, $P_{\text{O}_2} = 100$~mTorr, $t = 120$~min) STO substrates.}
  \label{fig:AFM_STO}
\end{figure}

\clearpage

\begin{figure}[h]
  \centering
  \includegraphics[width=5.5cm]{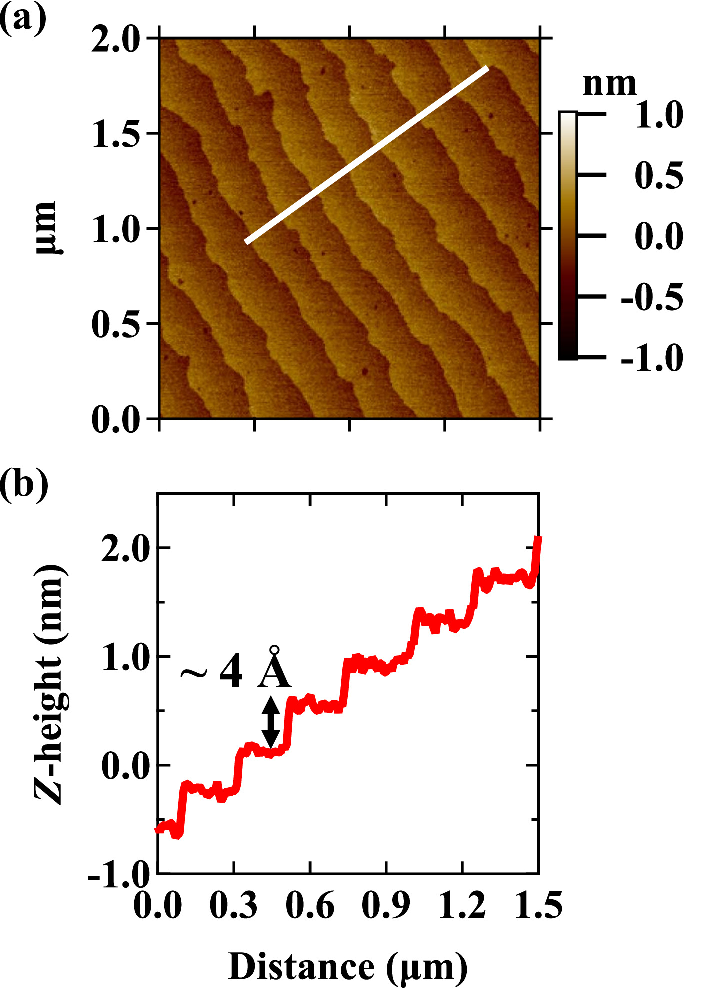}
  \caption{Surface morphology of 34~nm-thick-SRO thin films on a bare STO substrate ($T = 725^\circ$C, $P_{\text{O}_2} = 100$~mTorr, $t = 120$~min). (a)~AFM image and (b)~line profile along the white solid line after plane-fit leveling.}
  \label{fig:AFM_SRO_Lineprofile}
\end{figure}

\clearpage

\begin{figure*}[h]
  \centering
  \includegraphics[width=.9\textwidth]{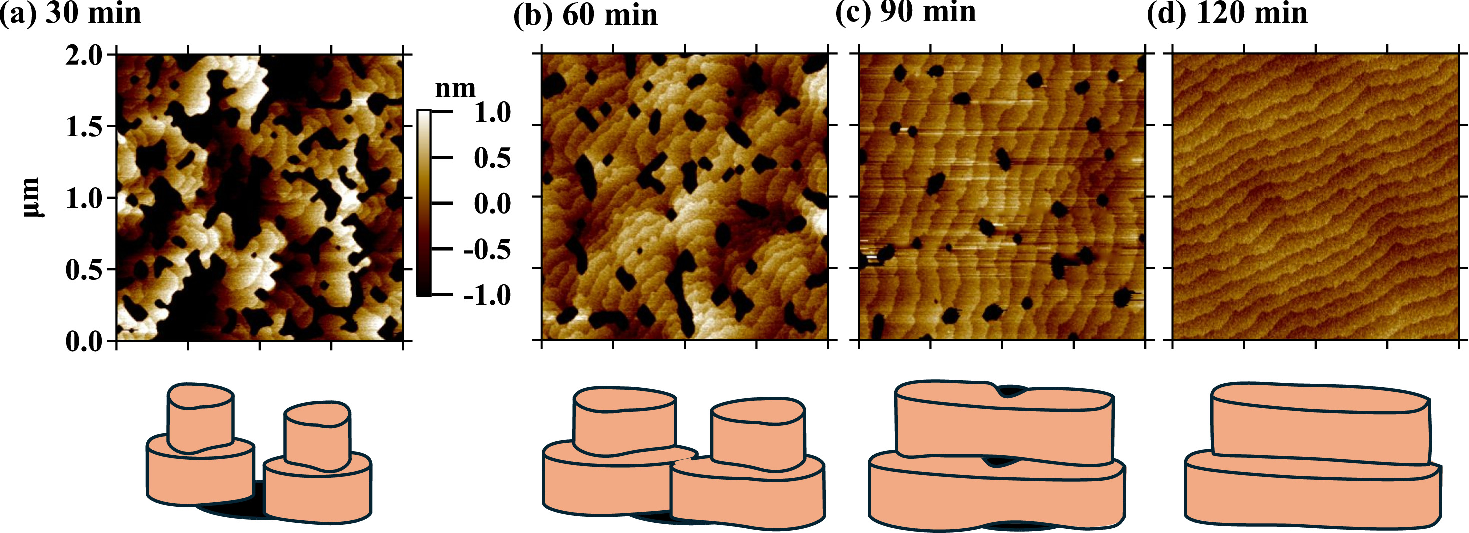}
  \caption{Surface morphology of SRO/STO ($T = 700^\circ$C, $P_{\text{O}_2} = 100$~mTorr, $d = 4.5$~cm) and growth schematics of the self-organized step-terrace structure at different growth times: (a)~$t = 30$~min, (b)~$60$~min, (c)~$90$~min, and (d)~$120$~min. The SRO film grown for 120~min has a thickness of approximately 34~nm. All AFM images are acquired over a $2~\mu\mathrm{m} \times 2~\mu\mathrm{m}$ scan area.}
  \label{fig:AFM_SRO_time_dependence}
\end{figure*}

\clearpage

\begin{figure*}[h]
  \centering
  \includegraphics[width=12cm]{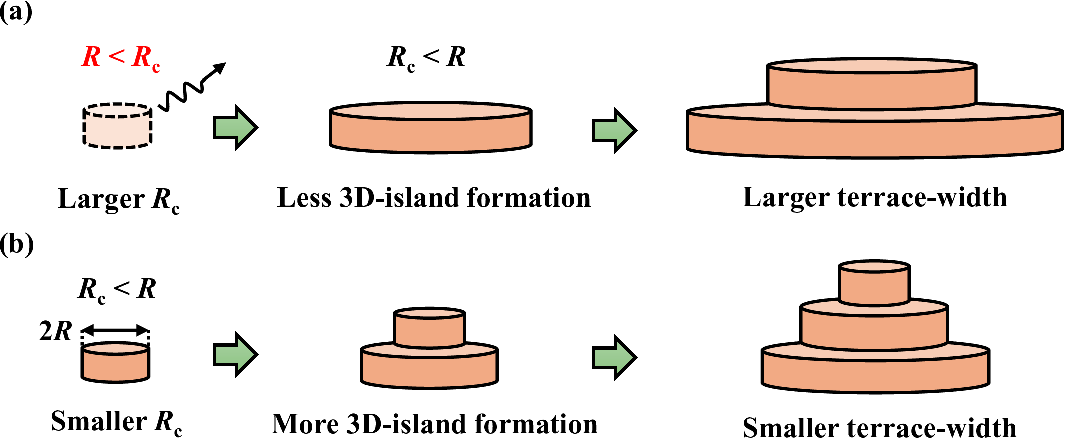}
  \caption{Schematic illustration of island growth mechanisms at the SRO/STO interface at (a)~$725^\circ$C and (b)~$700^\circ$C.}
  \label{fig:Schematics_Island_growth}
\end{figure*}

\clearpage

\begin{figure*}[h]
  \centering
  \includegraphics[width=14cm]{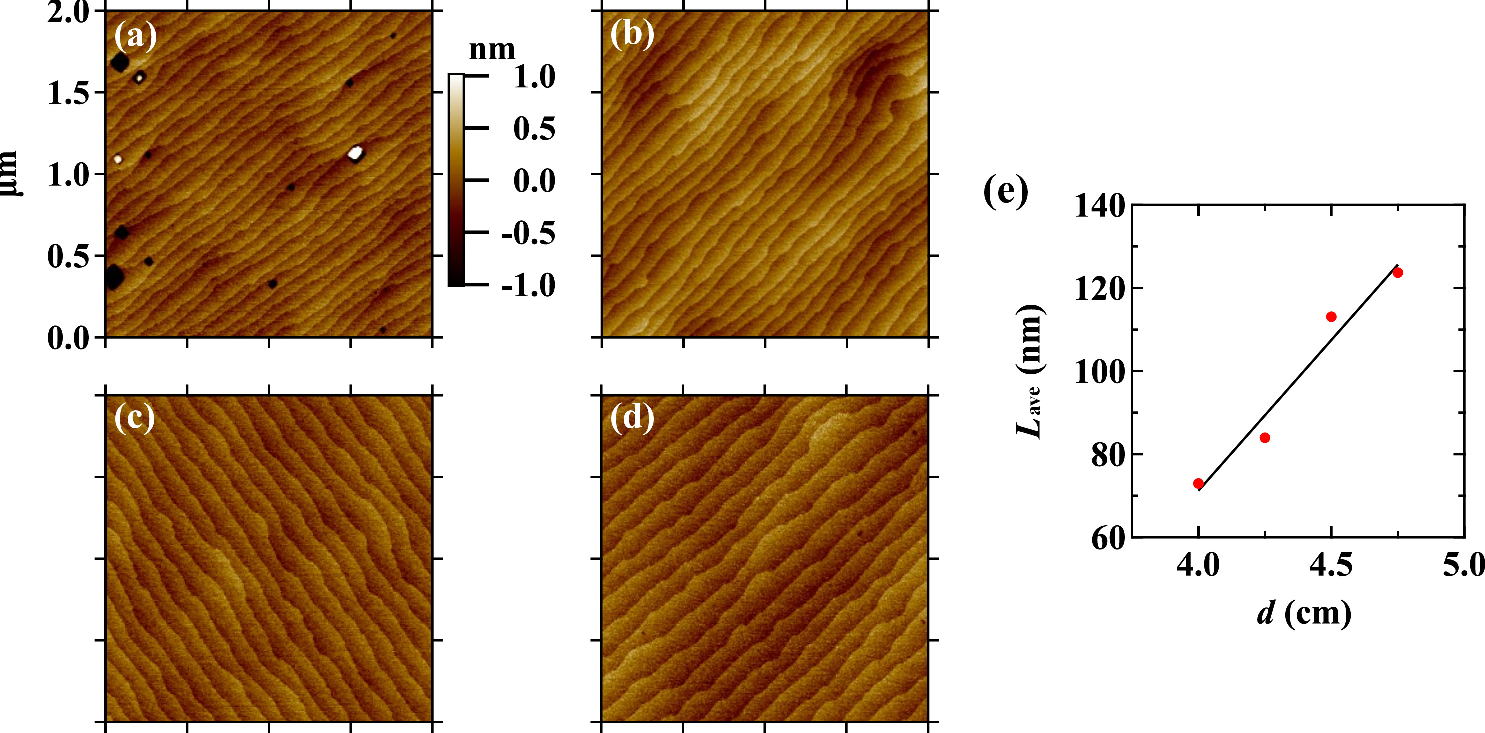}
  \caption{Surface morphology of SRO/STO ($T = 700^\circ$C, $P_{\text{O}_2} = 100$~mTorr, $t = 120$~min, thickness $\approx 34$~nm) for different target-substrate distances: (a)~$d = 4.0$~cm, (b)~$4.25$~cm, (c)~$4.5$~cm, and (d)~$4.75$~cm. (e)~Average terrace width $L_\text{ave}$ as a function of $d$. All AFM images are
acquired over a $2~\mu\mathrm{m} \times 2~\mu\mathrm{m}$ scan area.}
  \label{fig:AFM_SRO_dTS_dependence}
\end{figure*}

\clearpage

\begin{figure*}[h]
  \centering
  \includegraphics[width=10cm]{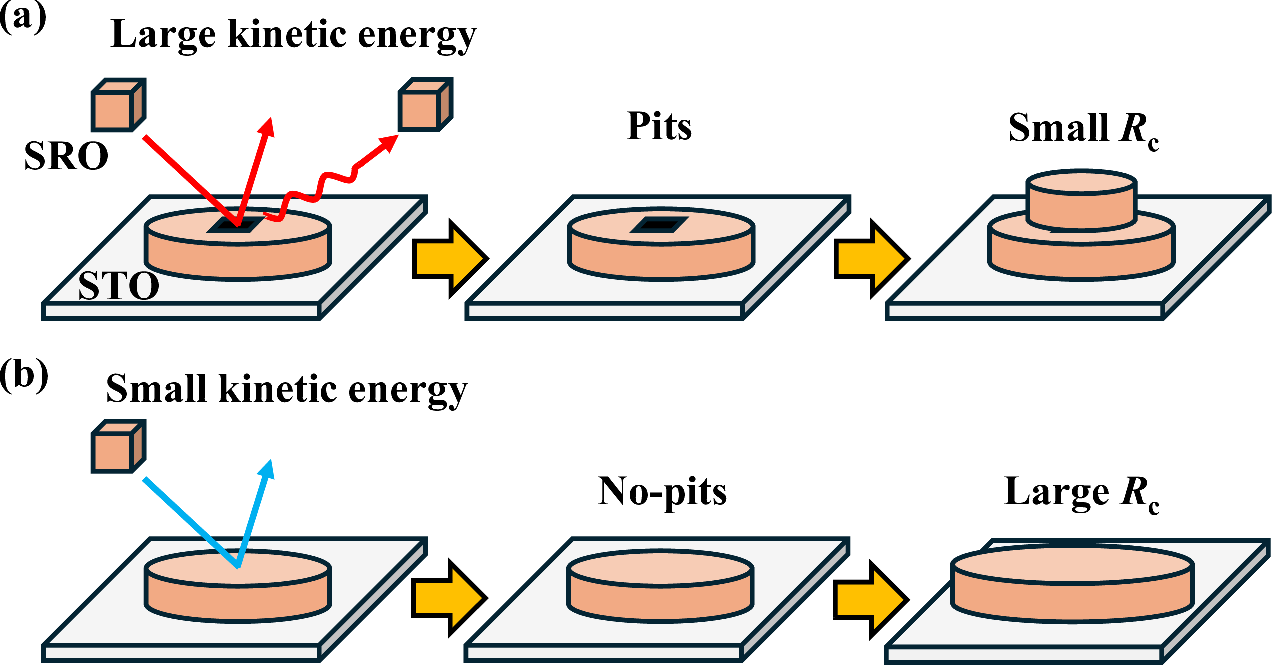}
  \caption{Schematic illustration of island growth mechanisms of SRO films with (a)~smaller and (b)~larger $d$.}
  \label{fig:Schematics_Island_growth_dTS}
\end{figure*}

\clearpage

\begin{figure*}[h]
  \centering
  \includegraphics[width=13cm]{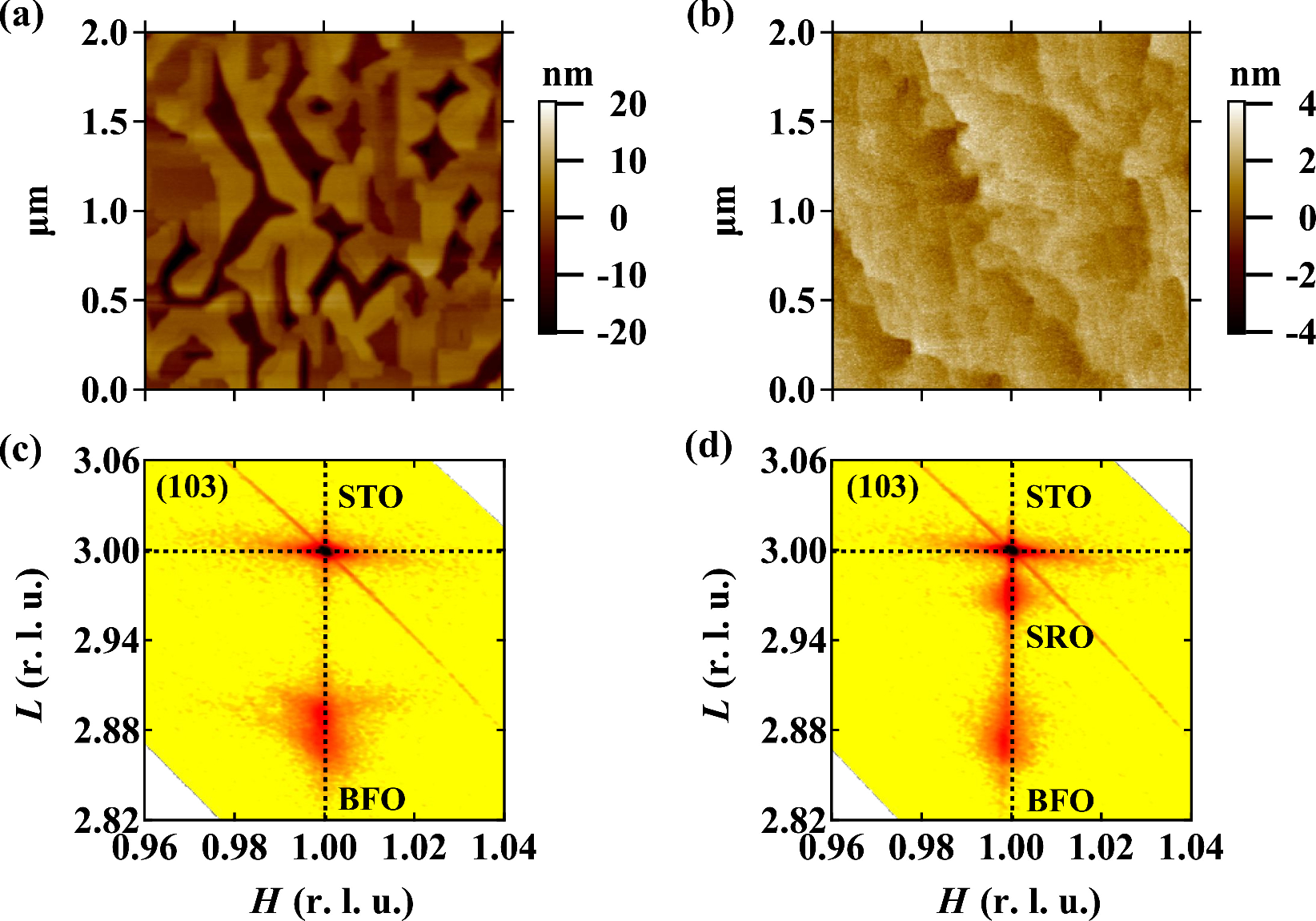}
  \caption{Surface morphology of 33~nm-thick-BFO thin films grown under identical conditions on (a)~the as-received STO substrate and (b)~the SRO/STO sample exhibiting a self-organized step-terrace structure. The corresponding root-mean-square roughness values are approximately 5.2~nm and 0.56~nm, respectively. (c) and (d)~show the RSM patterns for the respective samples.}
  \label{fig:AFM_RSM_BFO}
\end{figure*}

\clearpage
%=Supplementary Materials================================
\def\textpagefraction{.5}
\begin{center}
    {\Large\bfseries Supplementary Materials}
\end{center}
\setcounter{figure}{0}
\renewcommand{\thefigure}{S\arabic{figure}}

\section*{Growth of SRO/STO without using shadow-mask}
\begin{figure}[h]
  \centering
  \includegraphics[width=7cm]{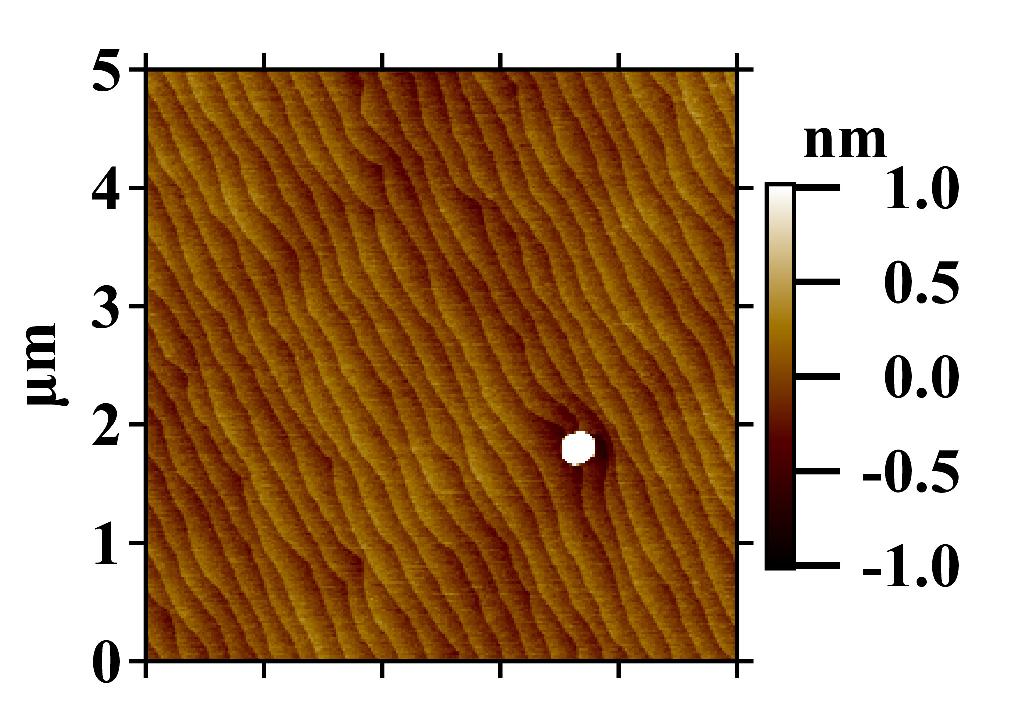}
  \caption{Surface morphology of 12.7~nm-thick-SRO deposited on mixed-terminated STO~(100) without using the 3D-shadow mask.}
  \label{fig:Supplementary_AFM_SRO_no_mask}
\end{figure}

We deposited SRO on a mixed-terminated STO (100) substrate without using 
a shadow mask in a separate PLD chamber under different growth conditions: 
$T = 650^\circ$C, $P_{\text{O}_2} = 400$~mTorr, $F = 4.2$~J/cm$^2$, 
$f = 10$~Hz, $d = 4$~cm, cooling under 600~Torr, $t = 15$~min 
(thickness $= 12.7$~nm). 
SRO was found to spontaneously form a step-terrace morphology even without 
the shadow mask, although droplets are also present 
(Fig.~\ref{fig:Supplementary_AFM_SRO_no_mask}). 
This confirms that the self-organized step-terrace formation is an 
intrinsic feature of SRO growth on mixed-terminated STO substrates, 
independent of the shadow-mask deposition. 
It is also noteworthy that the step-flow growth mode is established at a 
smaller film thickness than that observed in the shadow-mask experiments. 
Moreover, the average terrace width of approximately 185~nm is 
remarkably large compared to the values of 70--130~nm observed in the 
shadow-mask experiments (Fig.~5(e)), despite the lower substrate 
temperature of $650^\circ$C relative to the $700^\circ$C used in those 
experiments. 
These observations suggest that the relatively higher adatom diffusivity resulting from the absence of the shadow mask increases the critical island radius, promoting faster island coalescence and an earlier onset of step-flow growth at a smaller film thickness, as well as wider terraces after coalescence.

%===============================================================
%===============================================================
\section*{Influence of oxygen pressure on the SRO surface morphology}
\begin{figure}[h]
  \centering
  \includegraphics[width=16cm]{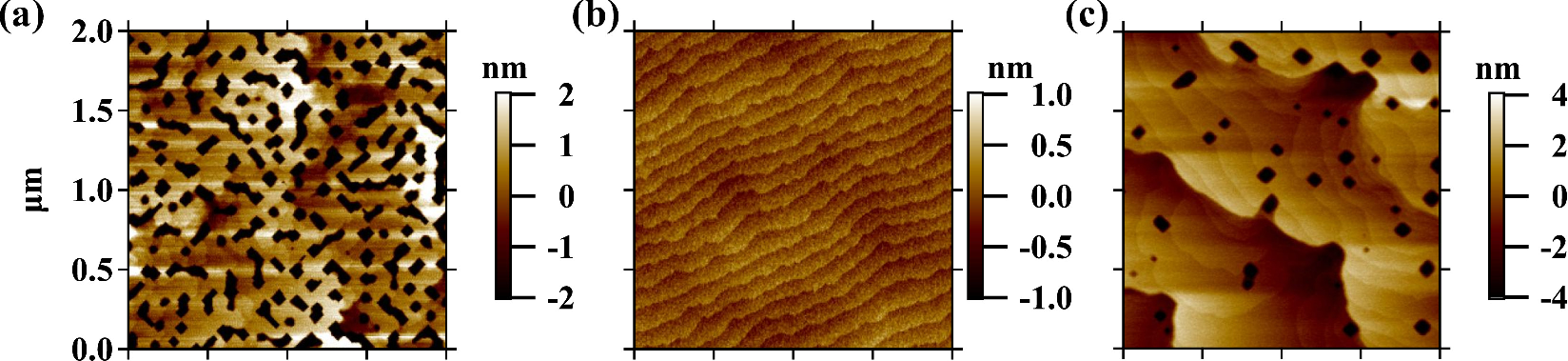}
  \caption{Surface morphology of SRO/STO grown under oxygen pressures of
(a) 80 mTorr, (b) 100 mTorr, and (c) 120 mTorr. All AFM images were
acquired over a $2~\mu\mathrm{m} \times 2~\mu\mathrm{m}$ scan area.
}
  \label{fig:Supplementary_AFM_PO2_dependence}
\end{figure}

The surface morphology of SRO thin films is strongly dependent on the oxygen pressure during growth (Fig.~\ref{fig:Supplementary_AFM_PO2_dependence}). A self-organized step-terrace structure forms only at the optimal oxygen pressure of 100~mTorr. At oxygen pressures of 80 or 120~mTorr, the surface becomes unstable and exhibits large pits, indicating a narrow optimal window for the step-flow growth mode. These results highlight the importance of precisely controlling the oxygen pressure.

%===============================================================
%===============================================================
\section*{Ferroelectric properties of BFO/SRO/STO}
\begin{figure}[h]
  \centering
  \includegraphics[width=16cm]{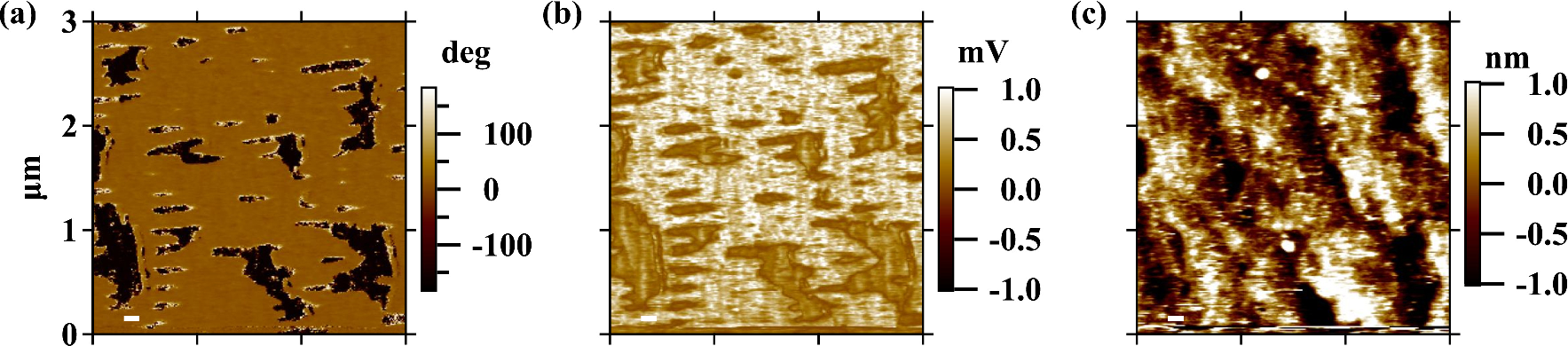}
  \caption{Lateral PFM images of BFO/SRO/STO: (a)~phase, (b)~amplitude, and (c)~$z$-height.  
All images were obtained over a $3~\mu\mathrm{m} \times 3~\mu\mathrm{m}$ scan area.
}
  \label{fig:Supplementary_PFM_BFO}
\end{figure}

The ferroelectric properties of BFO thin films deposited on SRO films 
with a self-organized structure were evaluated by piezoelectric force 
microscopy (PFM) in contact-resonance mode at room temperature using a 
Park Systems NX7. 
Prior to the PFM measurements, part of the BFO thin film was etched using 
KPZ-04 solution to expose the SRO layer as a bottom electrode. 
PFM phase, amplitude, and topographic ($z$-height) images were acquired as shown in Fig.~\ref{fig:Supplementary_PFM_BFO}.
Figures~\ref{fig:Supplementary_PFM_BFO}(a) and (b) show clear contrast 
in the lateral PFM phase and amplitude, whereas no contrast was observed 
in the vertical PFM phase signal (not shown). 
Since $71^\circ$ domains share the same out-of-plane polarization 
component, the vertical PFM signal is uniform across the film and thus 
shows no contrast between domains. 
The clear lateral PFM contrast therefore confirms the presence of a 
$71^\circ$ ferroelectric domain structure.

\clearpage

\clearpage

\end{document}